\documentclass[prl,nofootinbib,twocolumn,superscriptaddress,showpacs,showkeys]{revtex4-1}

\usepackage{graphicx,epsfig}
\usepackage{amsfonts,amsmath,amssymb,amsthm,amscd}
\usepackage{color}
\usepackage{dcolumn}
\usepackage{bm}
\usepackage{array}
\usepackage{float}
\usepackage[]{units}
\usepackage[colorlinks=true,linkcolor=blue,citecolor=blue]{hyperref}

\definecolor{teal}{rgb}{0, 0.5, 0.5}
\definecolor{darkmagenta}{rgb}{0.55, 0.0, 0.55}

\begin{document}

\title{Laser-particle collider for multi-GeV photon production}

\author{J. Magnusson}
\email{joel.magnusson@chalmers.se}
\affiliation{Department of Physics, Chalmers University of Technology, SE-41296 Gothenburg, Sweden}

\author{A. Gonoskov}
\affiliation{Department of Physics, Chalmers University of Technology, SE-41296 Gothenburg, Sweden}
\affiliation{Institute of Applied Physics, Russian Academy of Sciences, Nizhny Novgorod 603950, Russia}

\author{M. Marklund}
\affiliation{Department of Physics, Chalmers University of Technology, SE-41296 Gothenburg, Sweden}

\author{T. Zh. Esirkepov}
\affiliation{Kansai Photon Science Institute, National Institutes for Quantum and Radiological
Science and Technology (QST), 8-1-7 Umemidai, Kizugawa, Kyoto 619-0215, Japan}

\author{J. K. Koga}
\affiliation{Kansai Photon Science Institute, National Institutes for Quantum and Radiological
Science and Technology (QST), 8-1-7 Umemidai, Kizugawa, Kyoto 619-0215, Japan}

\author{K. Kondo}
\affiliation{Kansai Photon Science Institute, National Institutes for Quantum and Radiological
Science and Technology (QST), 8-1-7 Umemidai, Kizugawa, Kyoto 619-0215, Japan}

\author{M. Kando}
\affiliation{Kansai Photon Science Institute, National Institutes for Quantum and Radiological
Science and Technology (QST), 8-1-7 Umemidai, Kizugawa, Kyoto 619-0215, Japan}

\author{S. V. Bulanov}
\affiliation{Kansai Photon Science Institute, National Institutes for Quantum and Radiological
Science and Technology (QST), 8-1-7 Umemidai, Kizugawa, Kyoto 619-0215, Japan}
\affiliation{Institute of Physics ASCR, v.v.i. (FZU), ELI-Beamlines Project, 182 21 Prague, Czech Republic}
\affiliation{Prokhorov General Physics Institute RAS, Vavilov Str. 38 , Moscow 119991, Russia}

\author{G. Korn}
\affiliation{Institute of Physics ASCR, v.v.i. (FZU), ELI-Beamlines Project, 182 21 Prague, Czech Republic}

\author{S. S. Bulanov}
\affiliation{Lawrence Berkeley National Laboratory, Berkeley, California 94720, USA}

\begin{abstract}
As an alternative to Compton backscattering and bremsstrahlung, the process of colliding high-energy electron beams with strong laser fields can more efficiently provide both cleaner and brighter source of photons in the multi-GeV range for fundamental studies in nuclear and quark-gluon physics. In order to favor the emission of high-energy quanta and minimize their decay into electron-positron pairs the fields must not only be sufficiently strong, but also well localized. We here examine these aspects and develop the concept of a laser-particle collider tailored for high-energy photon generation. We show that the use of multiple colliding laser pulses with \unit[0.4]{PW} of total power is capable of converting more than \unit[18]{\%} of the initial multi-GeV electron beam energy into photons, each of which carries more than half of the electron energy.
\end{abstract}

\maketitle

The building and planning of several multi-PW laser facilities \cite{ELINP, ELI-beamlines, CORELS, XCELS, Vulcan} and the accessibility of PW-class systems \cite{danson.HPLSE.2015} have recently stimulated a strong interest in theoretical analysis of processes caused by the radiation reaction and by the phenomena of strong-field quantum electrodynamics (QED). The clarification of various theoretical aspects \cite{dipiazza.prl.2010,bulanov.nima.2011,dipiazza.rmp.2012, ilderton.prd.2013, ilderton.plb.2013,vranic.prl.2014,seipt.pra.2018,kharin.prl.2018} as well as the development of analytical \cite{gonoskov.pop.2018} and numerical \cite{nerush.prl.2011, duclous.ppcf.2011, sokolov.pop.2011, ridgers.jcp.2014, gonoskov.pre.2015, niel.pre.2018, derouillat.cpc.2018} approaches has been instrumental in revealing various peculiar effects such as stochasticity \cite{neits.prl.2013, green.prl.2014, esirkepov.pla.2015}, straggling \cite{shen.prl.1972, duclous.ppcf.2011}, quantum quenching \cite{harvey.prl.2017}, trapping in travelling \cite{zeldovich.spu.1975, li.prl.2014} and standing electromagnetic (EM) waves \cite{esirkepov.pla.2015,kirk.ppcf.2009,gonoskov.prl.2014,Jirka.pre.2016,kirk.ppcf.2016} and the alteration of ponderomotive effects \cite{esirkepov.pla.2015, fedotov.pra.2014}. These findings encouraged several promising proposals of both current \cite{cole.prx.2018, poder.prx.2018} and future experiments. This includes the creation of positron \cite{vranic.scirep.2018} and photon \cite{tamburini.scirep.2017, gonoskov.prx.2017, benedetti.natphot.2018, arefiev.prl.2016, wallin.pop.2017, lei.prl.2018,jansen.ppcf.2018} sources as well as probing fundamental aspects of QED and astrophysics by reaching extreme conditions through self-compression of laser-driven electron-positron plasmas \cite{efimenko.scirep.2018, efimenko.arxiv.2018}.

Apart from concepts of laser-based positron, gamma and X-ray sources \cite{esarey.pre.2002, rousse.prl.2004,chen.prl.2009,albert.ppcf.2014}, which may become favored through advances in laser wakefield acceleration (LWFA) \cite[and references therein]{leemans.prl.2014}, it is reasonable to also consider the use of optimally focused laser fields as targets for electron beams available with conventional accelerators. As it is today possible to create laser fields of sufficient strength for the emission of photons with energies comparable to that of the electrons, this process can provide an interesting alternative to Compton backscattering and bremsstrahlung, presently used in producing GeV-photons for probing nuclear and quark-gluon physics \cite{nedorezov.pu.2004}. In this letter we examine and develop the concept of such a laser-particle collider, applicable with both LWFA and conventional accelerators. In particular, we show that the use of a dipole focusing \cite{gonoskov.pra.2012, golla.epjd.2012, alber.jeos.2017} of multiple colliding laser pulses (MCLP) \cite{bulanov.prl.2010} makes sub-PW laser systems capable of converting more than \unit[18]{\%} of the initial electron beam energy into photons with more than half of the electron energy. This holds the potential for providing clean, ultra-bright sources of photons with energies ranging from a few to several tens of GeV for new fundamental studies. We note that the MCLP configuration was shown to significantly enhance a number of strong-field QED processes: from electron-positron pair production from vacuum \cite{bulanov.prl.2010,gonoskov.prl.2013}, to EM cascades \cite{gelfer.pra.2015,gonoskov.prx.2017,vranic.ppcf.2017, gong.pre.2017}.  

\textit{Motivating estimates --} An electron interacting with a strong laser field emits high-energy photons through the process of nonlinear Compton scattering. The probability of converting an electron with energy $\varepsilon_0 = \gamma mc^2$ into a photon carrying off a significant part of the electron's energy becomes large when the quantum nonlinearity parameter $\chi$ \cite{nikishov.jetp.1964} reaches values of the order of unity. Using a standing wave structure, e.g. through MCLP, provides a geometry that maximizes $\chi$ such that it can be estimated as $\gamma a_0/a_S$. We use relativistic units for both the laser field $a_0=eE/m\omega c$ and the Sauter-Schwinger field $a_S=eE_S/m\omega c \approx 4.1 \times 10^5 \lambda/\unit[1]{\mu m}$. Here $\hbar$ is Plank constant, $c$ is speed of light, $\omega$ and $\lambda$ are the laser frequency and wavelength, respectively, $e$ and $m$ are the charge and mass of the electron respectively. Efficient conversion therefore occurs for $a_0 \gtrsim a_S/\gamma$. If the electron experiences weaker fields, such that $\chi \ll 1$, but over an extended period of time its energy can be depleted through the emission of low-energy quanta ($\sim \chi \gamma mc^2$).  On the other hand, for sufficiently strong fields, a high-energy photon can in turn decay into an electron-positron pair, through multi-photon Breit-Wheeler, as it propagates through and interacts with the laser field for a sufficiently long time. 

\begin{figure}[t!]
	\includegraphics[width=\columnwidth]{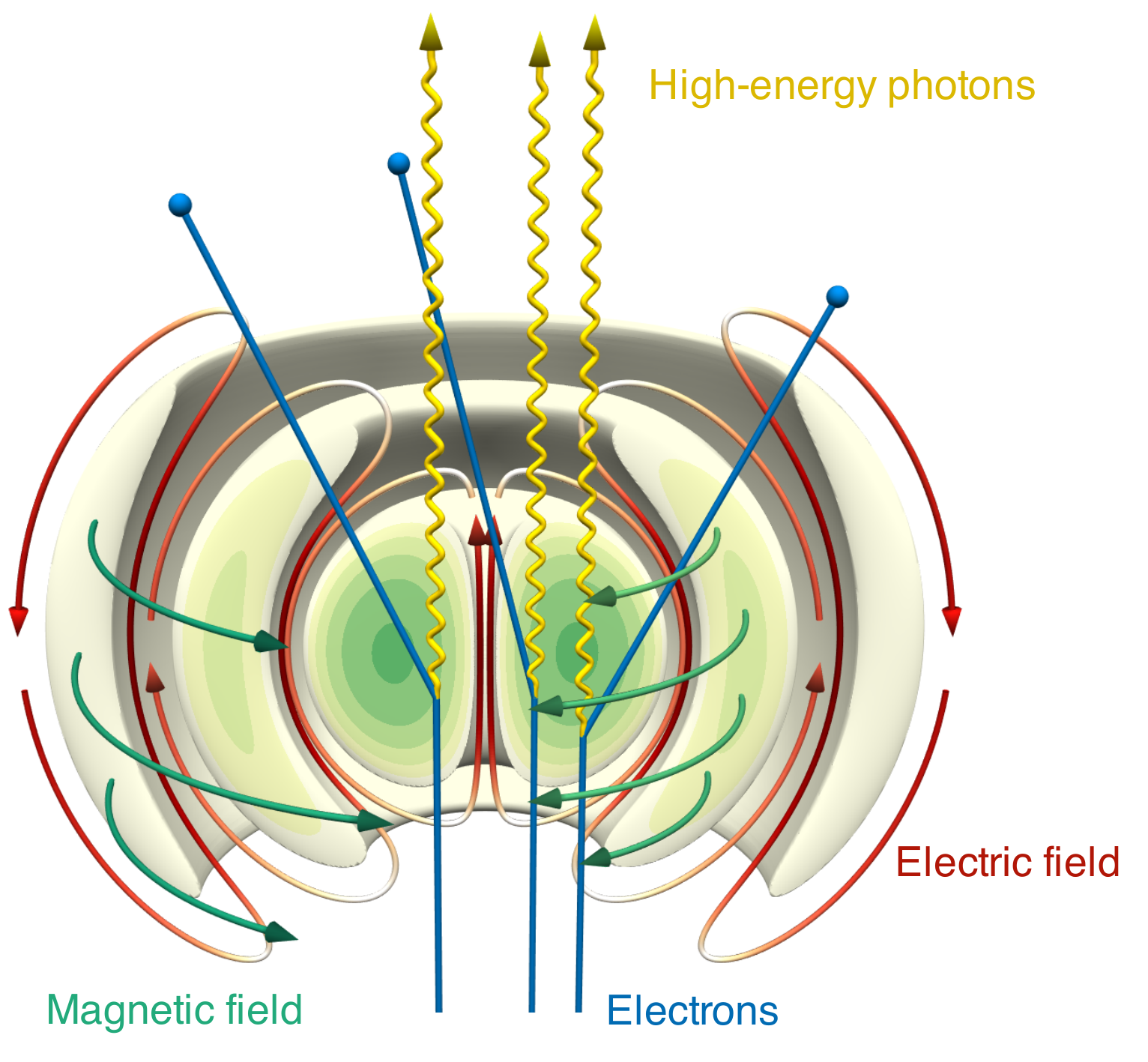}
	\caption{Conceptual visualization of the setup, where high-energy electrons (blue) are injected along the axis of an intense dipole wave. In this field, the electrons will emit large amounts of high-energy photons (yellow). The polarization of the shown field is that of an electric dipole, with a poloidal electric field (red) and a toroidal magnetic field (green).}
	\label{setup}
\end{figure} 

Using $\chi\gg 1$ approximation for the rates of these QED processes we can estimate the scale-lengths for both the photon generation $l_\mathrm{rad} \approx 15 \lambda_C \gamma^{1/3} (a_0/a_S)^{-2/3}$ and its decay $l_\mathrm{pair} \approx 3 l_\mathrm{rad}$, where $\lambda_C = \unit[2.43\times 10^{-10}]{cm}$ is the Compton wavelength and where we in the latter estimate assume that the photon energy is $\gamma mc^2/2$. As conversion efficiency, we here use the ratio $N_{1/2}/N$, where $N$ is the total number of electrons passing through the field and $N_{1/2}$ is the total number of photons with an energy above $\gamma mc^2/2$ and that escapes the interaction region. In order to maximize the yield according to this measure, one needs the field to be (1) sufficiently strong ($\chi \gtrsim 1$) for generating high-energy photons and (2) localized to within $\sim l_\mathrm{rad}$ in order to hamper the conversion of the generated photons into pairs by the same field. We see that $l_\mathrm{rad}$ increases with decreasing $a_0$ and can thus consider $a_0 = a_s/\gamma$ to be the minimal field amplitude that provides $\chi \sim 1$. For larger electron energies the radiation length increases as $l_\mathrm{rad} \sim 15 \lambda_C \gamma$ and for approximately $\unit[10]{GeV}$ it is on the order of the optical wavelength $\unit[1]{\mu m}$, which can be related to the diffraction limit for the field localization. 
For energies larger than this we can substitute $l_\mathrm{rad} \sim \unit[1]{\mu m}$ into the expression for $l_\mathrm{rad}$ and obtain an estimate for the optimal field amplitude $a_0 \approx 0.1 \gamma^{1/2}$.

\begin{figure}[t!]
	\includegraphics[width=\columnwidth]{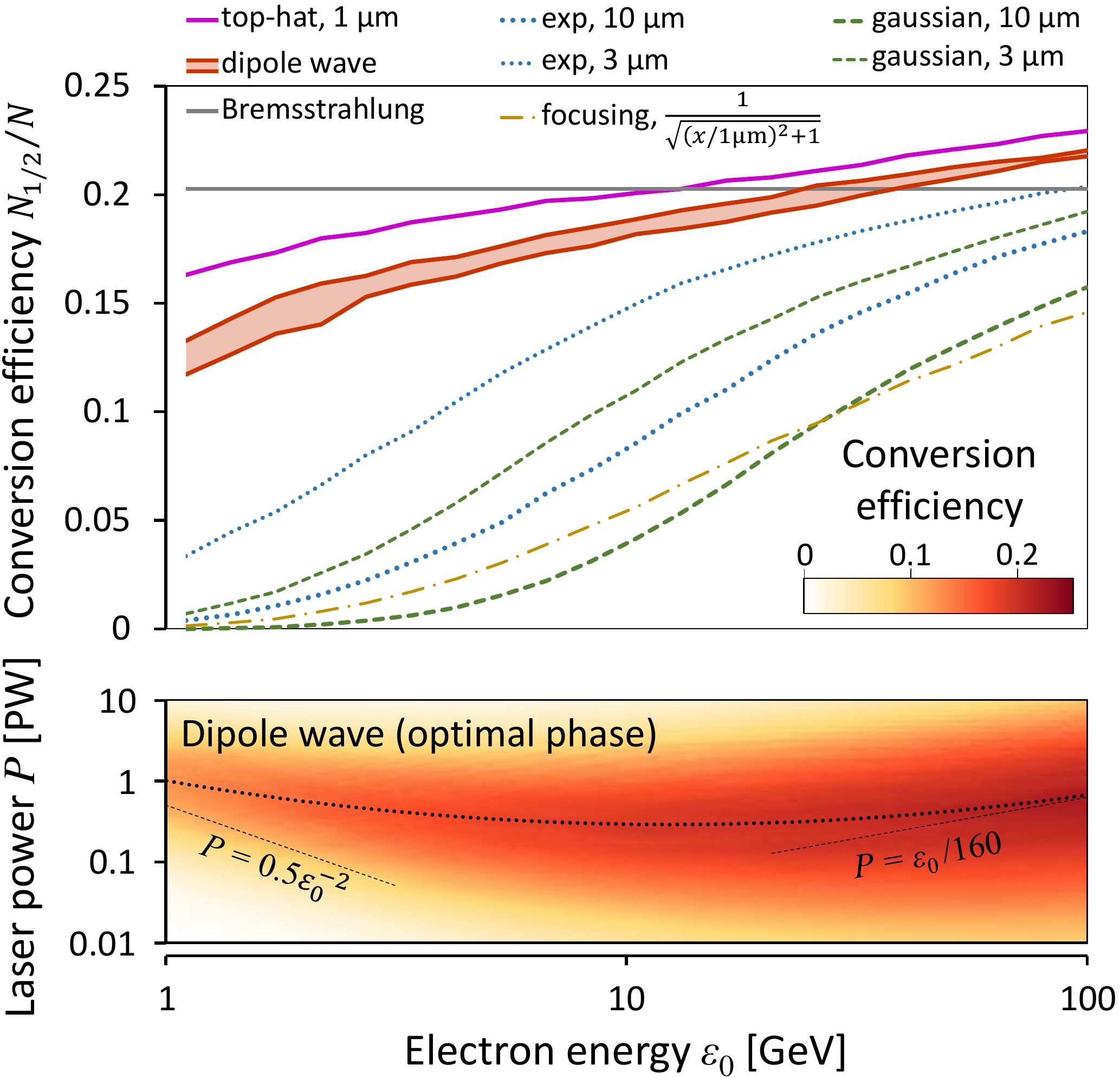}
	\caption{The computed conversion efficiency for optimal amplitude as a function of initial electron energy for various shapes of the field amplitude, $S(x)$, and scale-lenghts as shown in the legend above. The electrons and all generated particles are assumed to be ultra-relativistic (propagating at the speed of light) and thus experience fields given by $a_0/a_S = S(ct)$. We also show the maximal conversion efficency for breamsstrahlung, achieved at optimal thickness for arbitrary target material, see Ref.~\cite{blackburn.ppcf.2018}.
	The result for the dipole wave is shown with bounds corresponding to different phases. The conversion efficiency for the optimal phase is shown on the lower panel as a function of laser power $P$ and initial electron energy $\varepsilon_0$.}
	\label{comparison}
\end{figure}

We see that the scale-length of the field plays a crucial role in creating an efficient laser-particle collider for multi-GeV photon production. A remarkable field localization is provided by the $4\pi$-focusing geometry of a dipole wave \cite{gonoskov.pra.2012} (see fig.~\ref{setup}), which notably surpasses the diffraction limit. For a certain phase, an electron moving at a distance of $\lambda/3$ from and parallel to the dipole axis observes a field localized to within $0.3 \lambda$ (FWHM) and passes through the peak of the magnetic field at an amplitude of $\approx 500 \sqrt{P\,\left[\mathrm{PW}\right]}$, where $P$ is the total power of focused laser radiation. Using an averaged amplitude of $300\sqrt{P\,\left[\mathrm{PW}\right]}$ over the size of this field we can estimate the optimal power for efficient high-energy photon production following from the conditions formulated above for low ($\varepsilon_0 \ll \unit[10]{GeV}$) and high ($\varepsilon_0 \gg \unit[10]{GeV}$) electron energies, and determine numerically the intermediate value ($\varepsilon_0 \sim \unit[10]{GeV}$) from the data of Fig.~\ref{comparison}:
\begin{equation}\label{opt}
P^\mathrm{opt}(\varepsilon_0) \approx
\begin{cases}
0.5\left(\frac{\varepsilon_0}{\unit[1]{GeV}}\right)^{-2} \unit{PW}  & \text{if } \varepsilon_0 \ll \unit[10]{GeV} \\
\unit[0.4]{PW} & \text{if } \varepsilon_0 \sim \unit[10]{GeV} \\
\frac{\varepsilon_0}{\unit[160]{GeV}}\unit{PW} & \text{if } \varepsilon_0 \gg \unit[10]{GeV}
\end{cases}
\end{equation}
where $P^\mathrm{opt}(\varepsilon_0)$ is the power that maximizes $N_{1/2}/N$ for a given electron energy $\varepsilon_0$.

The absolute values of conversion efficiency in this dipole field is obtained numerically and shown in Fig.~\ref{comparison} together with the results for several other shapes of the field. In these computations we assume that both particles and photons propagate at the speed of light along a fixed direction, but with variable energy along the trajectory. The QED processes are modelled with the adaptive event generator described in Ref.~\cite{gonoskov.pre.2015}. The results are presented for optimal field amplitudes, determined individually for each field shape and initial electron energy. For the dipole field the variation due to the phase is presented as a band. In the lower panel we also show the conversion efficiency for the dipole wave as a function of both laser power $P$ and electron energy $\varepsilon_0$. One can see a reasonable agreement with the estimates (\ref{opt}), which we show with dashed lines. It is notable that fairly accessible PW systems are capable of reaching efficiencies as high as that of bremsstrahlung, while in addition admitting high concentrations of generated photons and a clean environment for experiments.

\textit{Simulations --} To assess further the properties of the proposed source we turn to large-scale simulations of the interaction process, where the fields and particle trajectories are more accurately modelled. The setup is investigated for laser powers and electron beam energies available currently or, in the near future, by using three dimensional particle-in-cell simulations. In this field geometry the achievable $\chi$ is well within the quantum regime, and so the numerical study is performed using the QED-PIC code \textsc{ELMIS} \cite{gonoskov.pre.2015}.

The simulations were carried out with a simulation box of $\unit[8]{{\mu}m}\times\unit[8]{{\mu}m}\times\unit[8]{{\mu}m}$ divided into $128 \times 128 \times 128$ cells (this spatial resolution is sufficient since we do not consider regimes of dense plasma formation and its dynamics). The dipole field is generated at the boundary of this region with a wavelength $\lambda = \unit[1]{{\mu}m}$ and cycle-averaged power $P$. This power is kept constant throughout the entire simulation in order to probe the interaction at a fixed supplied power, while allowing for a self-consistent field evolution that takes a potential suppression due to pair cascades into account, in the studied regime. 

Electrons were injected into the simulation box along the dipole axis of symmetry and with given energy $\varepsilon_0$. This beam of electrons was modelled as having a Gaussian spatial envelope, with a FWHM waist $w = \unit[1]{{\mu}m}$ and a FWHM length $L = \unit[5]{{\mu}m}$, corresponding to a duration of $\tau_0 = \unit[16.7]{fs}$. The total charge of the beam was $\unit[100]{pC}$, which translates into a total electron count of $N = 6.2\times 10^8$, and a peak density of $\unit[10^{20}]{cm^{-3}}$.

Statistics were gathered on all photons and positrons generated inside and leaving the primary interaction region and constitutes the main results of this study. Statistics is also gathered on all the electrons, both those injected and generated through Breit-Wheeler pair creation. 
In order to extend the earlier definition of generation efficiency, we calculate the total number of photons $N_x$ above a given energy threshold ($\varepsilon_\mathrm{th} = x\gamma mc^2$), for all photons escaping the interaction region.

\begin{figure}[t!]
\includegraphics[width=1\columnwidth]{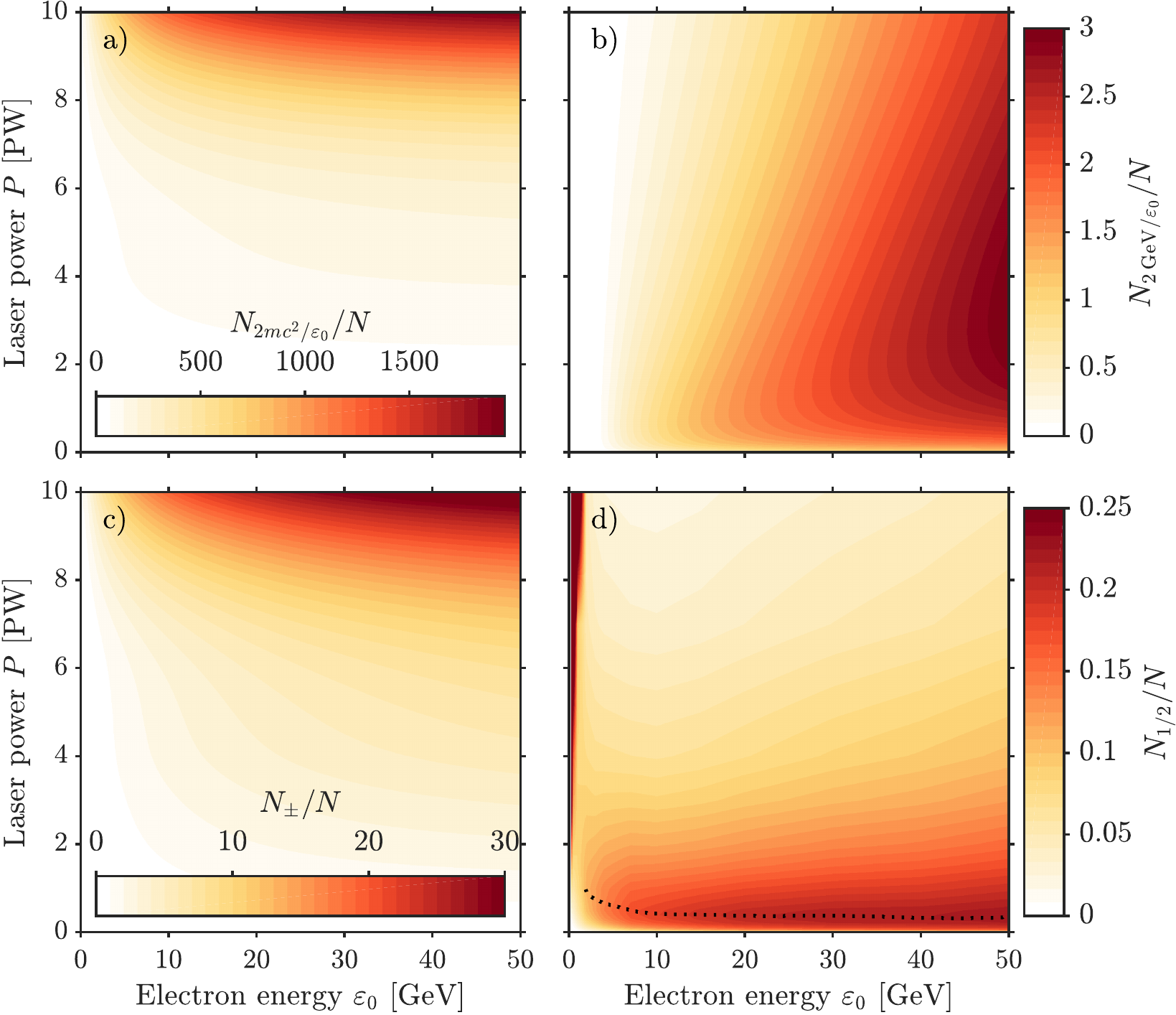}
\caption{Total number of photons detected above an energy threshold $\varepsilon_\mathrm{th}$ of (a) $2mc^2$, (b) $\unit[2]{GeV}$ and (d) $\varepsilon_0/2$, where $E$ is the electron beam energy. The values are normalized to the number of electrons in the beam ($N_{\varepsilon_\mathrm{th}/\varepsilon_0}/N$). (c) Total number of generated electron-positron pairs at the end of the simulation, also normalized to the number of incoming electrons ($N_\pm/N$).}
\label{fig:Nph}
\end{figure}

\textit{Results --} The photon generation efficiency is presented in Figure \ref{fig:Nph} for different cut-off energies and as a function of laser power and electron beam energy. It shows an intuitive trend for low cut-off energies (Figure \ref{fig:Nph}a), where both higher power and beam energy consistently translates into larger photon numbers, above the given threshold energy. However, for increasingly higher cut-offs (Figure \ref{fig:Nph}b) the efficiency instead displays an optimal laser power, for a given beam energy. This comes from the fact that as the laser power is increased, the pair production rate also increases. As a result, a smaller fraction of the high-energy photons escape the high-field region and instead fuel a shower-type cascade \cite{mironov.pra.2014}. 

\begin{figure*}[t!]
	\includegraphics[width=1\textwidth]{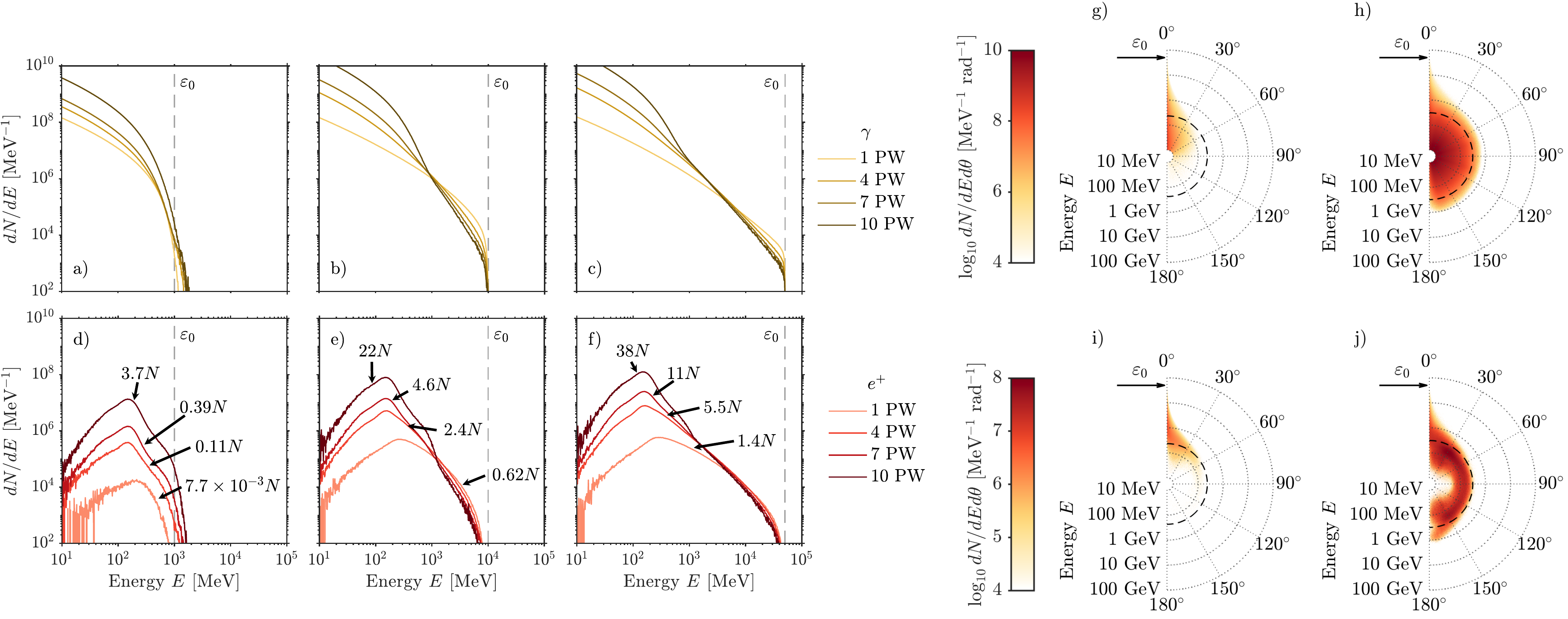}
	\caption{Comparison between energy spectra of (a-c) photons (yellow) and (d-f) positrons (red) for different laser powers. Indicated with a dashed line is the initial electron energy: (a, d) \unit[1]{GeV}, (b, e) \unit[10]{GeV}, (c, f) \unit[50]{GeV}. The total number of positrons is also indicated in (d-f), where $N$ is the total initial number of electrons in the beam. The energy-angle distribution of the generated (g-h) photons and (i-j) electrons are shown for an electron beam energy $\varepsilon_0$ of \unit[50]{GeV} and a laser power $P$ of (g, i) \unit[1]{PW} and (h, j) \unit[10]{PW}.}
	\label{fig:spectra_comp}
\end{figure*}

In Figure \ref{fig:Nph}d we show the efficiency for generating photons above half the initial electron beam energy. 
The high efficiency region at $\varepsilon_0 \lesssim \unit[1]{GeV}$ is due to reacceleration in the laser field, which makes multiple emission of these photons possible. Furthermore, there is also a region of high efficiency at large electron beam energies. As could already be seen from the two high-cutoff figures, in this region the efficiency initially increases with increasing laser power, but eventually drops off as the photon decay into pairs becomes dominant. The generation efficiency is here seen to be optimal around $\unit[0.4]{PW}$, and with an electron beam energy of $\unit[10]{GeV}$ it is possible to reach an efficiency of $\unit[18]{\%}$.

To further elucidate the interplay between the shower cascade and the suppression of high energy photons with increasing laser powers, it is informative to compare the photon and positron spectra for different laser powers and beam energies (Fig.~\ref{fig:spectra_comp}). Here it can be clearly seen that the number of photons above $\unit[1]{GeV}$ is strongly suppressed for high laser powers, while the number of generated pairs increases, leading to the photon spectra for these laser powers to almost coincide for energies $>\unit[1]{GeV}$.

The total number of pairs produced is similarly shown in Figure \ref{fig:Nph}c, again normalized to the number of electrons in the beam. This shows a clear monotonic increase in the number of generated pairs for both increasing beam energy and laser power, as expected. For sufficiently large values we have a cascade of pairs being produced due to non-linear Breit-Wheeler. This region is also separated from both regions of high-energy photon production (compare Fig. \ref{fig:Nph}c and \ref{fig:Nph}d). 

In Fig. \ref{fig:spectra_comp} we show energy-angle distribution of electrons and photons as they leave the interaction region, for a $\unit[50]{GeV}$ e-beam and either $\unit[1]{PW}$ or $\unit[10]{PW}$ of laser power. We note that some electrons still travel in the initial beam direction, and those electrons have the highest final energy. All other electrons, scattered by the EM field in all directions have much lower energies, limited by a several hundred MeV threshold. This can be explained by the fact that these electrons are moving in the radiation dominated regime, where the emission of photons dominate the electron dynamics. If we consider a strong rotating electric field, then the electron energy in such a field is given by $(a_0/\varepsilon_{rad})^{1/4}$ \cite{bulanov.pre.2011}, where $\varepsilon_{rad}=4\pi r_e/3\lambda$ and $r_e$ is the classical electron radius. This estimate works reasonably well for the case of a dipole wave, predicting maximum energies similar to that obtained in simulations.

\textit{Conclusions --}
In summary, we have investigated the interaction of a highly energetic electron beam with an intense laser pulse, in a geometry of optimal focusing, and assessed its capabilities as a source of GeV-level photons. We find that in this geometry and for large initial electron energies, increasing the laser power above $\unit[1]{PW}$ leads to an increasingly stronger shower cascade, hampering the yield of high-energy photons. To efficiently generate photons above a few GeV we find that there is an optimal laser power of around $\unit[0.4]{PW}$, around which it is possible reach efficiencies in excess of $\unit[18]{\%}$.

Under such conditions, it would therefore be possible to use a significant amount of the power available to $\unit[10]{PW}$-class systems to generate high-energy electron beams, having to dedicate only a smaller fraction to the photon generation. It also means that even with an imperfect geometry, it may still be possible to reach this regime with currently available laser powers, by compensating for the imperfections with a larger supplied power than suggested here.

Furthermore, our results show that improving the capabilities of current and future laser systems in terms of realizing more complex field geometries and synchronization, parallel to the current efforts of increasing the available laser power, may be a worthwhile investment.

\begin{acknowledgments}
\textit{Acknowledgments --}
SSB acknowledges support from the Office of Science of the US DOE under Contract No. DE-AC02-05CH11231.
JKK acknowledges support from JSPS KAKENHI Grant Number 16K05639. 
SVB acknowledges support at the ELI-BL by the project High Field Initiative (CZ.02.1.01/0.0/0.0/15 003/0000449) from European Regional Development Fund.
The research is partly supported by the Russian Science Foundation Project No. 16-12-10486 (A.G., computations), by the Swedish Research Council grants No. 2013-4248, 2016-03329 (M.M.) and 2017-05148 (A.G.), and by the Knut \& Alice Wallenberg Foundation (A.G., J.M., M.M.). The simulations were performed on resources provided by the Swedish National Infrastructure for Computing (SNIC) at HPC2N.
\end{acknowledgments}


\bibliography{Laser-particle_collider}{}
\bibliographystyle{aipnum4-1}

\vfill

\end{document}